\documentclass[11pt]{article}
\usepackage[utf8]{inputenc}
\usepackage{authblk}
\usepackage{mathtools}
\usepackage{amsmath}
\usepackage{amssymb}
\usepackage{caption}
\usepackage{abstract}
\usepackage{enumitem}
\usepackage[breaklinks]{hyperref}
\usepackage[superscript]{cite}
\hypersetup{colorlinks=true, allcolors=blue}

\usepackage{xcolor}

\usepackage{setspace}

\usepackage{lmodern}

\usepackage[left=3cm, right=3cm, top=3cm]{geometry}

\newcommand{\tsb}[1]{\textsf{\textbf{#1}}}

\vspace{-10pt}
\title{\huge\textsf{An open-source, end-to-end workflow for multidimensional photoemission spectroscopy}}

\author[1,*]{R. Patrick Xian}
\author[2]{Yves Acremann}
\author[3]{Steinn Y. Agustsson}
\author[1]{Maciej Dendzik}
\author[2]{Kevin B\"uhlmann}
\author[4]{Davide Curcio}
\author[5]{Dmytro Kutnyakhov}
\author[5,6]{Federico Pressacco}
\author[5]{Michael Heber}
\author[1]{Shuo Dong}
\author[1]{Tommaso Pincelli}
\author[3]{Jure Demsar}
\author[5,6,$\dagger$]{Wilfried Wurth}
\author[4]{Philip Hofmann}
\author[1]{Martin Wolf}
\author[1,7]{Markus Scheidgen}
\author[1,*]{Laurenz Rettig}
\author[1,*]{Ralph Ernstorfer}

\affil[1]{Fritz Haber Institute of the Max Planck Society, 14195 Berlin, Germany}
\affil[2]{Laboratory for Solid State Physics, ETH
Zurich, 8093 Zurich, Switzerland}
\affil[3]{Department of Physics, University of Mainz, 55128 Mainz, Germany}
\affil[4]{Department of Physics and Astronomy, Interdisciplinary Nanoscience Center (iNANO), Aarhus University, 8000 Aarhus C, Denmark}
\affil[5]{DESY Photon Science, 22607 Hamburg, Germany}
\affil[6]{Department of Physics, University of Hamburg, 22761 Hamburg, Germany}
\affil[7]{Department of Physics, Humboldt University of Berlin, 12489 Berlin, Germany}
\affil[$\dagger$]{Deceased}
\affil[*]{Corresponding authors: \textsf{xian@fhi-berlin.mpg.de}, \textsf{rettig@fhi-berlin.mpg.de}, \textsf{ernstorfer@fhi-berlin.mpg.de}}

\begin{document}

\date{}
\maketitle

\begin{quote}
\begin{center}
    \section*{Abstract}
\end{center}
\textsf{Characterization of the electronic band structure of solid state materials is routinely performed using photoemission spectroscopy. Recent advancements in short-wavelength light sources and electron detectors give rise to multidimensional photoemission spectroscopy, allowing parallel measurements of the electron spectral function simultaneously in energy, two momentum components and additional physical parameters with single-event detection capability. Efficient processing of the photoelectron event streams at a rate of up to tens of megabytes per second will enable rapid band mapping for materials characterization. We describe an open-source workflow that allows user interaction with billion-count single-electron events in photoemission band mapping experiments, compatible with beamlines at $3^{\text{rd}}$ and $4^{\text{th}}$ generation light sources and table-top laser-based setups. The workflow offers an end-to-end recipe from distributed operations on single-event data to structured formats for downstream scientific tasks and storage to materials science database integration. Both the workflow and processed data can be archived for reuse, providing the infrastructure for documenting the provenance and lineage of photoemission data for future high-throughput experiments.}
\end{quote}

\section*{Introduction}
Many disciplines in the natural sciences are increasingly dealing with densely sampled multidimensional datasets. The scientific workflows to obtain and process them are becoming increasingly complex due to the provenance and structure of the data and the information needed to be extracted and analyzed \cite{Pruneau2017,Deelman2018}. In materials science and condensed matter physics, various spectroscopic and structural characterization techniques produce experimental data of distinct formats and characteristics. Their creation and understanding require customized processing and analysis pipelines designed by specialists in the respective fields. The growing incentive for building experimental materials science databases \cite{Zakutayev2018} that complement established theoretical counterparts \cite{Himanen2019} calls for open-source and reusable workflows for data processing \cite{Pizzi2018,Perkel2019} that transform raw data to shareable formats for downstream query, analysis and comparison by non-specialists of the experimental techniques \cite{Hill2016,Draxl2018}. Among the various properties associated with materials, the electronic band structure (EBS) of condensed matter systems is of vital importance to the understanding of their electronic properties in and out of equilibrium. Multidimensional photoemission spectroscopy (MPES) \cite{Schonhense2015,Medjanik2017,Schonhense2018} is an emerging technique that bears the potential of high-throughput EBS characterization through band mapping experiments and holds promise as an enabling technology for building experimental EBS databases, where data integration requires traceable knowledge of the processing steps between the archived and the raw format. Here we present an open-source workflow that focuses on band mapping data from MPES. In the following, we briefly introduce the technology of MPES and the associated data processing, before providing details on the workflow from raw data to database integration.
\begin{figure}[htb!]
  \begin{center}
    \includegraphics[scale=0.85]{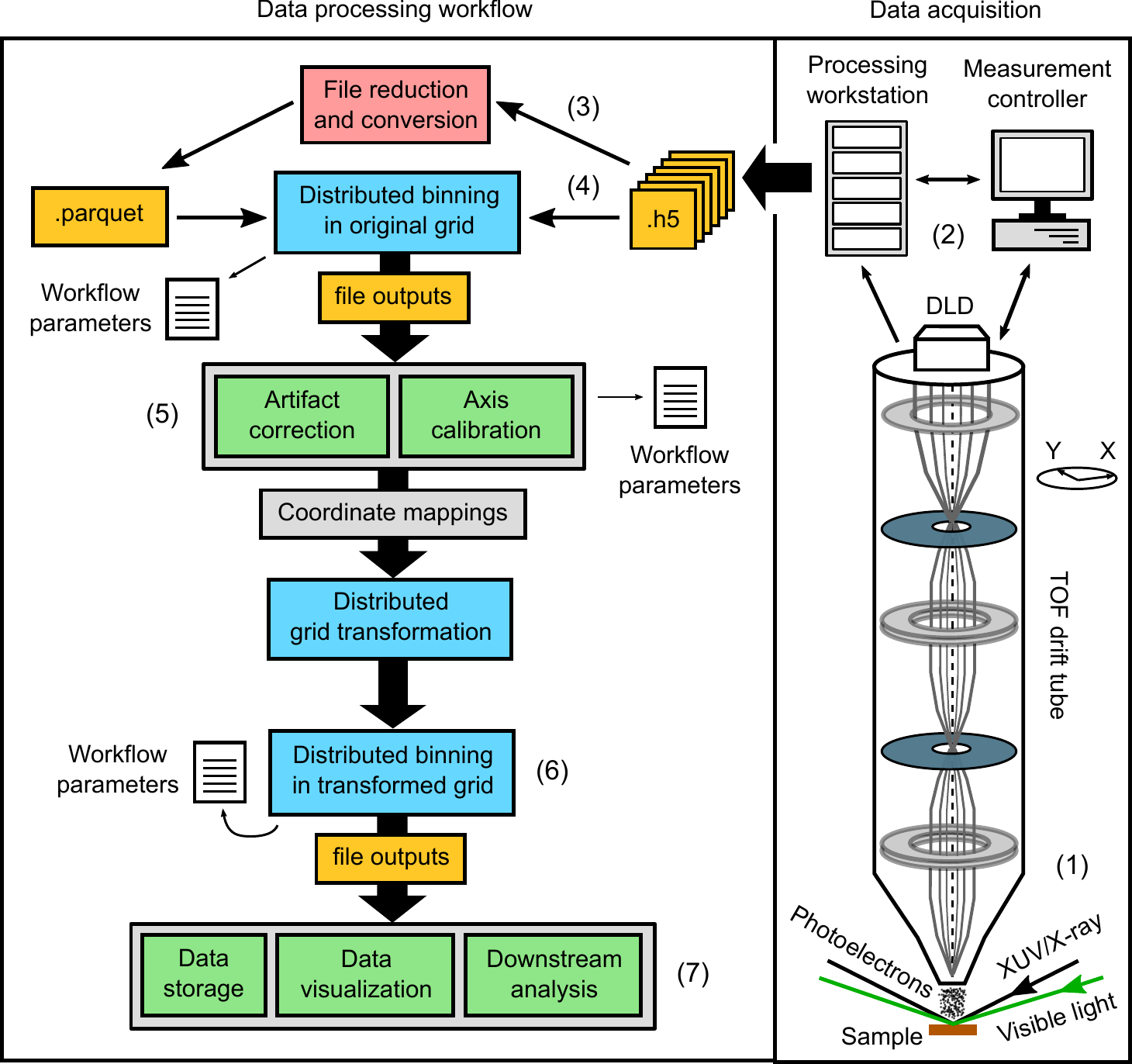}
    \caption{\textbf{Schematic of the workflow in MPES.} The data acquisition in MPES starts from (1) photoelectrons liberated by the extreme UV (XUV) or X-ray photons travelling through the lens systems and the TOF tube to trigger detection events on the delay-line detector (DLD). (2) Single-event data acquisition is monitored and controlled by the measurement controller computer. The raw data are first streamed and stored onto a hard drive in HDF5 format (.h5) and subsequently processed in the workflow through (3) file reduction (optional), (4)(6) distributed binning, (5) artifact correction and axis calibrations, carried out at the single-event or the binned data levels. At the end of the workflow, other data formats are generated (such as HDF5, MAT or TIFF) for (7) storage, visualization or downstream analysis for extracting relevant physical parameters. Critical parameters within the workflow may be exported (as workflow parameters files), shared and reused for processing other datasets.}
    \label{fig:workflow}
  \end{center}
\end{figure}

MPES, also called momentum microscopy (MM), is born out of the recent integration of time-of-flight (TOF) electron spectrometers with delay-line detectors (DLDs) and improved electron-optic lens designs \cite{Kromker2008,Ovsyannikov2013,Damm2015,Tusche2015}. Compared with the earlier generations of angle-resolved photoemission spectroscopy (ARPES) \cite{Damascelli2003,Yang2018} using hemispherical analyzers to measure the 2D energy-momentum distribution of the photoemitted electrons \cite{Suga2014}, MPES is capable of recording single-electron events simultaneously sorted into the $(k_x, k_y, E)$ coordinates ($E$: electron energy, $k_x,k_y$: parallel momentum components) in band mapping experiments, obviating the need for scanning across sample orientations and subsequent data merging as is the case for similar experiments using a single hemispherical analyzer. Operation of the TOF DLD in MPES requires a pulsed photon source and is directly compatible with 3$^{\text{rd}}$ and 4$^{\text{th}}$ generation light sources \cite{Couprie2014} as well as laboratory-based table-top setups \cite{Chiang2015,Puppin2018,Corder2018,Buss2019}, harnessing their high repetition rates in the range of multi-kilohertz to megahertz to drastically improve the detection speed and efficiency. Mapping of the 3D band structure with sufficient signal-to-noise ratio (SNR) may be achieved on the timescale of minutes. The technological convergence opens up the possibilities to record 3D datasets in dependence of one or more additional parameters, such as spatial location $I(x, y, k_x,k_y,E)$, probe photon energy, $I(k_x,k_y,E,k_z)$ \cite{Medjanik2017}, spin-polarization, $I(k_x,k_y,E,S)$ \cite{Schonhense2015}, or pump-probe time in time-resolved MM, $I(k_x,k_y,E,t)$ \cite{Kutnyakhov2020} within a reasonable time frame.

From the data perspective, the pulsed sources with high repetition rates generate densely sampled data at rates of multiple megabytes per second (MB/s), which has brought about challenges in data processing and management compared with conventional ARPES experiments. The raw data in MPES are single photoelectron events registered by the DLD and the physical quantities related to the detected events are streamed in parallel to the storage files in a hierarchical file format (e.g. HDF5 \cite{Folk2011}). A typical dataset involves $10^7-10^{10}$ detected events with a total size of up to a few hundred of gigabytes (GBs), depending on the number of coordinates measured (3D or 4D) and the required SNR. Unlike the large 2D or 3D image-based datasets, such as those obtained in various forms of optical \cite{Weiler2014,Ker2018} and electron microscopy techniques \cite{Levin2016,Aversa2018}, processing and conversion of tabulated single-event data requires additional steps of statistical computing for conversion into standard images. This motivates the current workflow development for efficient data processing and analysis. In data processing and calibration, experiments performed at different facilities share similar procedures going from the raw events to the multidimensional hypervolume with calibrated axes, which is the basis for archiving and downstream analysis. To maintain reproducibility for the particular data source characteristics, data structure and processing procedure, we have summarized the workflow (see Fig. \ref{fig:workflow}) into two open-source software packages (\textsf{hextof-processor} \cite{hextof-processor} and \textsf{mpes} \cite{mpes}), with similar design principles for coping with large-scale facility and table-top experiments, respectively. The core of our approach includes distributed statistical processing at the single-event level using parameters calibrated and determined from preprocessed volumetric datasets, which enables effective instrument diagnostics, artifact correction, and sample condition monitoring. The algorithms involved balance physical knowledge and existing methods in image processing and computer vision. The workflow is illustrated next with data obtained at some of the electron momentum microscopes currently in operation, such as the HEXTOF (high energy X-ray time-of-flight) measurement system \cite{Kutnyakhov2020} at the free-electron laser source FLASH \cite{Ackermann2007} at DESY, and the table-top high harmonic generation-based setup at the Fritz Haber Institute (FHI) \cite{Puppin2018} involving a commercial TOF and DLD (METIS 1000, SPECS GmbH). We use the material example of tungsten diselenide (WSe$_2$) measured at both experimental setups to demonstrate the workflow execution, because in momentum space, the patent features of WSe$_2$ band structure and the nonequilibrium dynamics initiated by optical excitation of WSe$_2$ have been thoroughly studied in the past (see Methods) \cite{Riley2014,Shallenberger2018,Bertoni2016,Dendzik2020,Kutnyakhov2020}. We expect the workflow described here to serve as a blueprint for upcoming software platforms in similar setups to be installed in other facilities or laboratories worldwide.

\section*{Results}
\begin{enumerate}[leftmargin=0pt, labelindent=0pt]
\item[] \textbf{Workflow description.} The workflow schematic shown in Fig.~\ref{fig:workflow} starts with raw single-event data from measurements. The data are (i) binned in a distributed fashion in the measurement coordinates, including each of the photoelectrons' position on the detector $(X,Y)$, its TOF, a digital encoder (ENC) axis, and others, if more than four dimensions are acquired in parallel. The binned histogram is (ii) used to estimate the numerical transforms for distortion correction and axis calibration. Next, these transforms are (iii) applied to the raw single-event data to convert the measurement coordinates to the physical axes, $(k_x, k_y, E, t_{\text{pp}})$ and others for higher dimensions (see also Fig. \ref{fig:workflow_mapping}). Finally, the single-event data are (iv) binned in the transformed grid to yield 3D, 3D+t or other higher-dimensional data with the correct axis values. The outcome may be exported in different formats for storage, visualization and downstream analysis.

\item[] \textbf{Tasks and software infrastructure.} Processing billion-count single-event data requires user interaction for data checking and distributed processing to reduce the time consumption. The general tasks in the workflow include the transformation of data streams to multidimensional histograms, artifact correction and axis calibration. These operations can be efficiently decomposed into column-wise operations of the distributed dataframe format offered by the \textsf{dask} package \cite{DaskDevelopmentTeam2016} in Python. While the use of \textsf{dask} dataframes provide the common foundation for interactivity with single events of \textsf{hextof-processor} and \textsf{mpes}, they distinguish themselves by the experimental requirements.
\begin{figure}[htbp!]
  \begin{center}
    \includegraphics[scale=0.5]{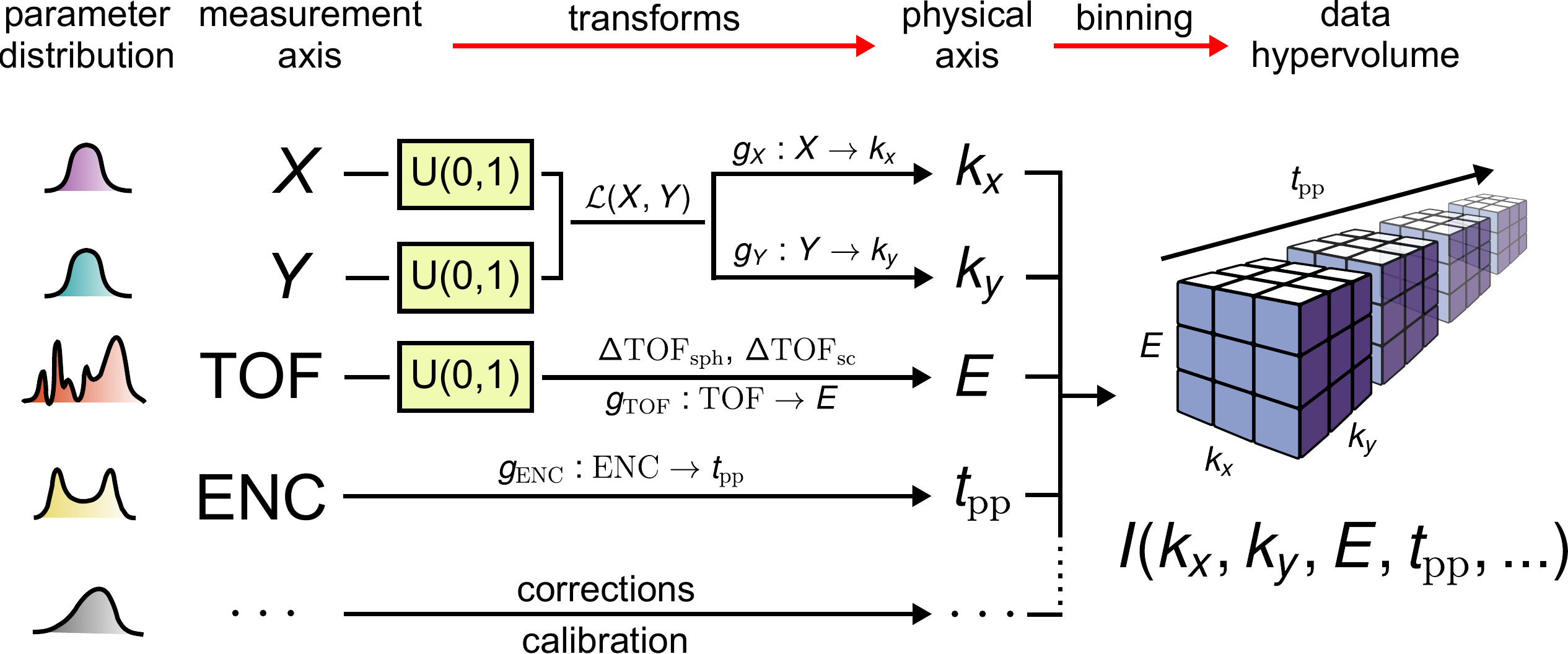}
    \caption{\textbf{Examples of workflow components.} Illustrations are given for artifact correction and axis calibration. Characteristic 1D distributions of the measured $X$, $Y$, TOF, ENC and an arbitrary axis are shown on the very left. $U(0, 1)$ represent uniformly distributed random noise added to suppress digitization artifacts (jittering or dithering). The transforms ($g$'s) are calibration functions that convert the values in the measurement axes to the physical ones. The transform $\mathcal{L}(X, Y)$ corrects the symmetry distortion, while the spherical timing aberration and space charge are compensated for by $\Delta\text{TOF}_{\text{sph}}$ and $\Delta\text{TOF}_{\text{sc}}$, respectively. Binning of the corrected single-event data over the calibrated physical axes yields a multidimensional hypervolume (right picture) of photoemission intensity data along with the physical axes values.}
    \label{fig:workflow_mapping}
  \end{center}
\end{figure}
At large-scale facilities, experiments often record a large number of machine parameters that need to be stored, though only a small number of relevant parameters are needed for downstream processing. Therefore, the \textsf{hextof-processor} package includes a parameter sampling step to retrieve intermediate tabulated data in the Apache Parquet format (\href{https://parquet.apache.org/}{https://parquet.apache.org/}), a column-based data structure optimized for computational efficiency. This approach reduces the processing overhead in searching through the raw data files every time when data are queried during the subsequent processing. As an open-source project, other beamtime-specific functionalities are added by users to the existing framework at every new experimental run. The \textsf{mpes} package adapts to the much simpler file structure produced at table-top experimental setups and makes direct use of the HDF5 raw data. It comes with added functionalities motivated by the existing issues encountered in data acquisition and downstream processing such as axis calibration, masking, alignment and different forms of artifact correction. The softwares come with detailed documentation and examples online for users to gain familiarity (see Code Availability).

\item[] \textbf{Artifact correction.} Artifacts in MPES data come from mechanical imperfections, stray fields (electric and magnetic), uncertainties in the alignment of the sample, light beams and the multistage electron-optic lens systems as well as the data digitization process. Minimizing and correcting instrumental imperfections plays an important role in the validity of downstream analysis. We carry out artifact correction sequentially at the level of single photoelectron events or the data hypervolume obtained from multidimensional histogramming (see Fig. \ref{fig:workflow_mapping}). The outcomes are illustrated using the correction of (1) digitization artifact (see Fig. \ref{fig:jitter}) and (2) spherical timing aberration artifact (see Fig. \ref{fig:spherabb}), with technical details in Methods.
\begin{figure}[htb!]
  \begin{center}
    \includegraphics[width=\textwidth]{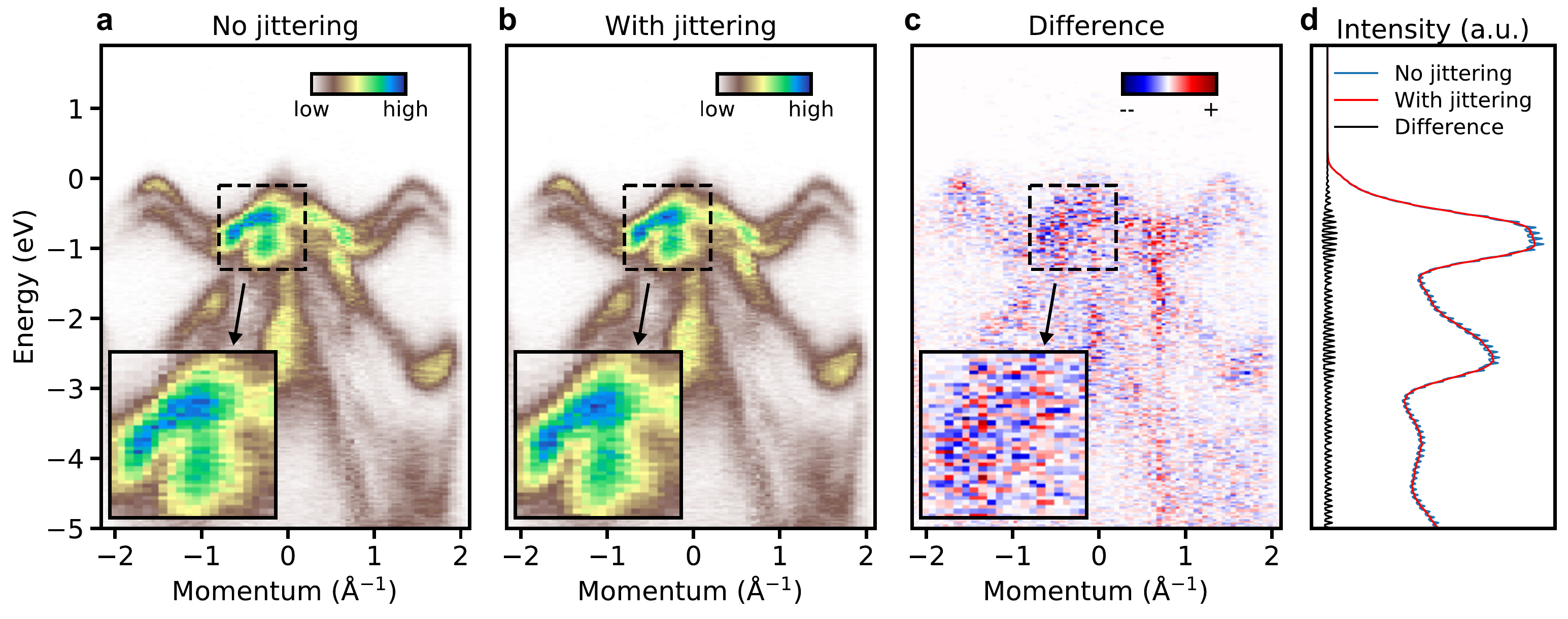}
    \caption{\textbf{Digitization artifact correction by histogram jittering.} Removal of the digitization artifact is illustrated with a 2D $k$-$E$ slice across the Brillouin zone center (at 0 $\mathrm{\AA}^{-1}$ momentum) of the band mapping dataset measured at FHI on WSe$_2$. The images before and after histogram jittering and their difference are shown in \tsb{a}, \tsb{b} and \tsb{c} respectively. A zoomed-in section of the data are shown in the insets in \tsb{a}-\tsb{c}. The effective removal of the digitization artifact is further demonstrated in the momentum-integrated energy distribution curves in \tsb{d}. The traces in \tsb{d} are computed by averaging horizontally over their corresponding 2D images in \tsb{a}-\tsb{c}.}
    \label{fig:jitter}
  \end{center}
\end{figure}

\item[] \textbf{Axis calibration.} To transform the measurement axes of the DLD into physically relevant axes for electronic band mapping, calibrations are required, as shown in Fig. \ref{fig:workflow_mapping}. The calibration functions are constructed with parameters derived from comparing physical knowledge of the materials (e.g. Brillouin zone size, Fermi level position) with the corresponding scales in data. They are applied either to the binned data hypervolume, or to the single-electron events raw data individually in a distributed fashion before binning. Details on the calibration data transforms are provided in Methods.
\begin{figure}[htb!]
  \begin{center}
    \includegraphics[scale=0.8]{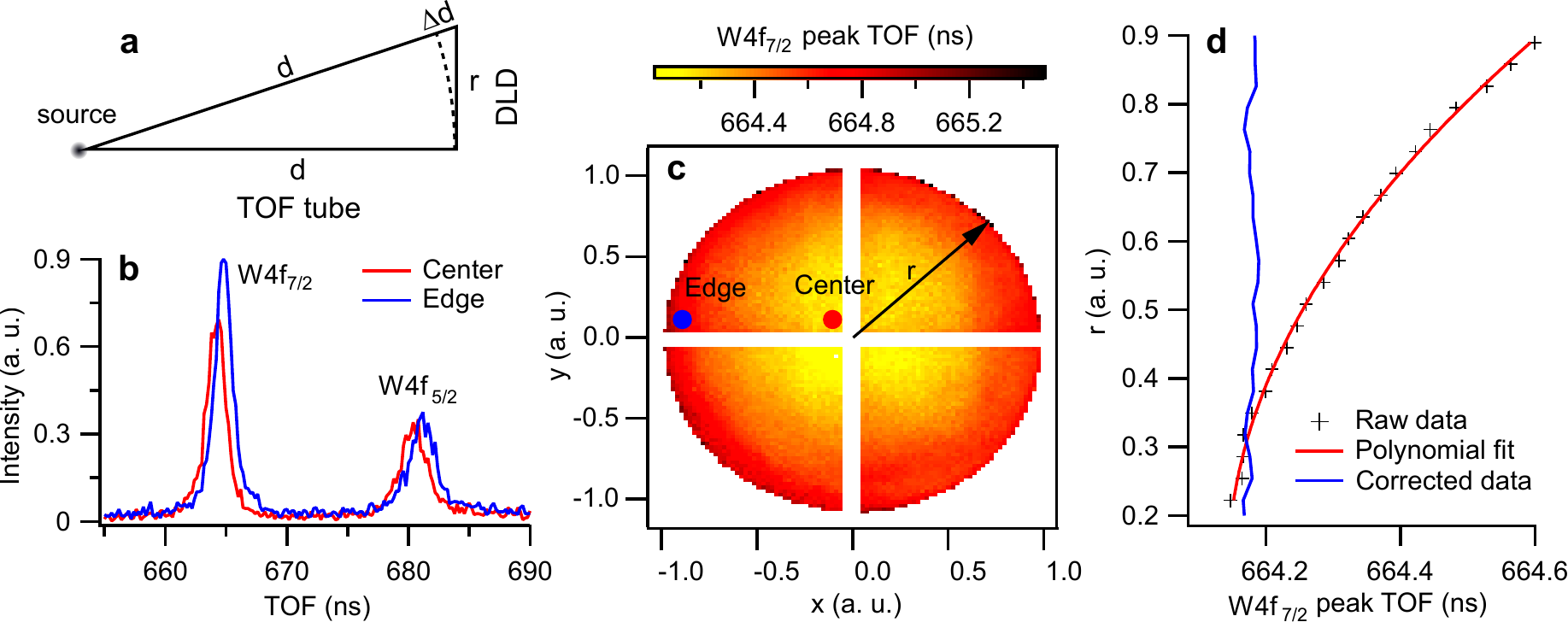}
    \caption{\textbf{Spherical timing aberration correction.} The correction is demonstrated using W4f core-level data measured at FLASH. The energy spacing between the W4f$_{7/2}$ and W4f$_{5/2}$ levels is about 2.1 eV \cite{Shallenberger2018}. \tsb{a.} Illustration of the geometric origin of the spherical timing aberration in the time-of-flight (TOF) drift tube. \tsb{b.} Comparison of the W4f spectra at the center and on the edge of the detector plane. The energy spectra are extracted from the corresponding regions, marked by the dots in the same blue and red colors, respectively, in \tsb{c}. The white stripes crossing at the detector center block the exposed edges of the four-quadrant detector quadrants. \tsb{d.} The uncorrected and corrected radial-averaged peak TOF positions for the W4f$_{7/2}$ core level.}
    \label{fig:spherabb}
  \end{center}
\end{figure}

\item[] \textbf{Data storage and format.} The simplistic form of the output data hypervolume derived from single-electron events includes non-negative scalar values of the photoemission intensity and the calibrated real-valued axes coordinates, including $k_x$, $k_y$, $E$, and other parameter dependencies such as the pump-probe time delay $t_{pp}$. These values are exported as HDF5, MAT or TIFF, with the metadata included as attributes of the files.

\item[] \textbf{Workflow archiving and reuse.} Computational workflows are valued by their the reproducibility \cite{Stodden2016}. Archiving and sharing the workflow parameters among users of the beamlines or facilities allow comparison between experimental runs and reuse for the simultaneous benefits of machine diagnostics and experimental efficiency. To achieve this, we store critical parameters generated within the workflow in a separate file as \textit{workflow parameters} (see Fig. \ref{fig:workflow}) during each step, including the numerical values used in binning, the intermediate parameters and coefficients of the correction and calibration functions, etc. They can be reused when loading into the processing of other datasets.
\begin{figure}[htb!]
  \begin{center}
    \includegraphics[scale=1]{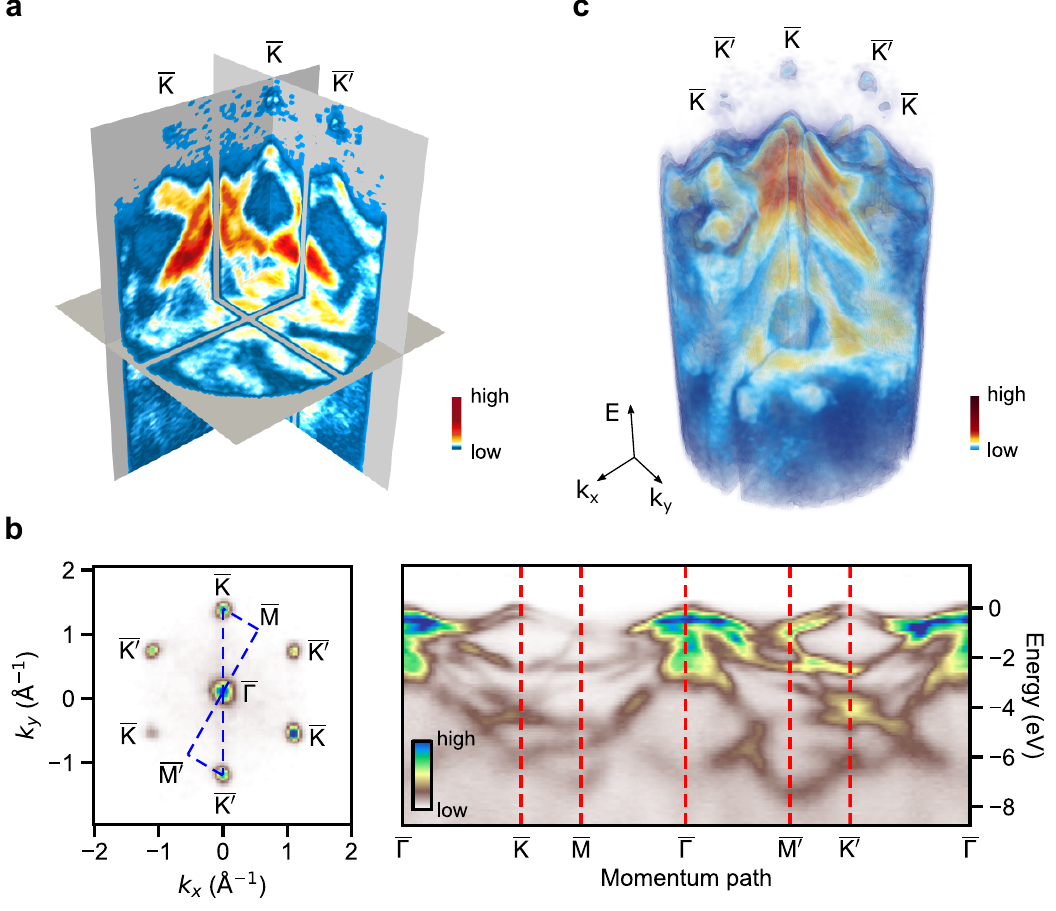}
    \caption{\textbf{Typical visual representations of the volumetric band mapping data.} The examples are illustrated using band mapping data of the layered semiconductor WSe$_2$, measured with the HEXTOF instrument at FLASH (\tsb{a},\tsb{c}) and the METIS detector at the FHI (\tsb{b}). The visualizations are \tsb{a.} the orthoslice representation, \tsb{b.} the band-path diagram (right) with the momentum path labelled in dashed blue line in the momentum $k_x$-$k_y$ plane (left), and \tsb{c.} the cut-out view. All color scales represent photoemission intensity. The letters label the high symmetry points of the hexagonal Brillouin zone of WSe$_2$ \cite{Riley2014}. The intensity features on the upper side nearby the $\overline{\mathrm{K}}$ and $\overline{\mathrm{K}^{\prime}}$ points result from nonequilibrium electronic populations following photoexcitation of electrons to the conduction band, while the intensities below are from valence band photoemission (see Methods for experimental conditions).}
    \label{fig:datarepr}
  \end{center}
\end{figure}

\item[] \textbf{Data visualization.} The adaptation of established scientific visualization methods in the physical sciences \cite{Hansen2005,Lipsa2012} to band mapping data should incorporate the requirements and knowledge of the data characteristics in this field of research. The band mapping data in 3D (multi-megavoxel) and 3D+t (multi-gigavoxel) include the inherent symmetries from the electronic band structure of the material, but the intensity modulations in the photoemission process \cite{Moser2017}, dynamics and sample condition disrupt the original symmetry. The overall goal is to emphasize the features of interest while exploiting the symmetry to simplify the visualization (see Methods). The output files from the processing pipeline are compatible with open-source visualization software such as \textsf{matplotlib} \cite{Hunter:2007}, \textsf{ParaView} \cite{Hansen2005} and \textsf{Blender} \cite{Blender}.

\item[] \textbf{Downstream analysis integration.} Typical photoemission data analysis involves extracting electronic band structure parameters, physical coupling constants and lifetimes via fitting of lineshapes \cite{Damascelli2003} or dynamical models \cite{Weinelt2002}, often carried out specific to the material under study. At the end of our distributed workflow, the data size is on the order of a few to tens of gigabytes, which can be directly loaded into memory on users' local machines for downstream data analysis with custom routines.

\item[] \textbf{Experimental metadata.} The metadata of the data files have a tree structure and contain information of the experimental setting, parameters of the pulsed light source, the detector and the sample under study. A list of top-level metadata parameters is presented in Table \ref{tab:meta}. A full and current list of all metadata parameters, including the top-level parameters and their constituent lower-level parameters, along with their definitions, units and values, is provided in Supplementary Tables 1-4. For database integration, an accompanying data parser (\textsf{parser-mpes}, see Code Availability) for MPES data has been written in accordance with existing standards \cite{Ghiringhelli2017} for computational materials science in NOMAD \cite{Draxl2018}, featuring an electronic version of the metadata parameter list in the file \href{https://gitlab.mpcdf.mpg.de/rpx/parser-mpes/-/blob/master/mpesparser/mpes.nomadmetainfo.json}{mpes.nomadmetainfo.json} online. The metadata parameter list and the data parser are versioned and are updated based on the corresponding changes in the data structure for photoemission spectroscopy experiments. The existing WSe$_2$ photoemission data have been integrated into the experimental section of the materials science database NOMAD (see Data Availability).
\begin{table}[htb!]
\centering
\caption{Top-level metadata parameters}
\begin{tabular}{|c|c|}
    \hline
    Category name & Description \\
    \hline
    General parameters & Descriptive information of the experiment and facility \\
    \hline
    Source parameters & Technical parameters relating to the photon source \\
    \hline
    Detector parameters & Technical parameters relating to the photoelectron detector \\
    \hline
    Sample parameters & Parameters relating to the material sample in experiment \\
    \hline
\end{tabular}
\label{tab:meta}
\end{table}
\end{enumerate}

\section*{Discussion}
We have designed and implemented an open-source, end-to-end workflow for processing single-event data produced in multidimensional photoemission spectroscopy, linking to downstream tasks, providing guidelines and software for integrating processed data into the NOMAD experimental materials science database. The distributed processing takes full advantage of the single-event data streams directly accessible from the TOF delay-line detector for event-wise correction and calibration and converts the raw events to the calibrated data hypervolume for project-specific downstream analysis. The functionalities within the workflow are publicly accessible through the software packages we have developed (\textsf{hextof-processor} \cite{hextof-processor} and \textsf{mpes} \cite{mpes}). The processing workflow is archived at each step of operation and the processed data may be integrated into experimental database with user-specified metadata. The methods described here are applicable to all existing types of multidimensional photoemission band mapping measurements beyond the static and time-dependent settings described here.

Our end-to-end workflow from raw data to processed data to database integration provides a fast-track and all-in-one solution to the demands for open experimental data and reproducible research in the materials science community \cite{Hill2016,Draxl2018}. The public repositories for the software packages are the foundations for phased future extension and integration with existing analytical tools in the photoemission spectroscopy community. The modular structure of the packages introduced here allows targeted upgrades by both temporary and dedicated maintainers and users. Casting the workflow in the Python programming environment provides the foundation for convenient incorporation of existing image processing and machine-learning resources \cite{Butler2018} for further exploration and understanding of the band mapping datasets, which contain rich information owing to the complex nature of the photoemission process \cite{Damascelli2003,Suga2014}. This is especially beneficial for broader adoption of photoemission since the interpretation of photoemission data is often linked to the observed or extracted outstanding features such as local intensity extrema, dispersion kinks and satellites, lineshape parameters and pattern symmetry \cite{Damascelli2003}, therefore, the access to experimental data and the potential integration with existing electronic structure-related software \cite{Pizzi2018,Ong2013,HjorthLarsen2017,MGanose2018} will facilitate method developments and the direct comparison between experimental results and theoretical band structure calculations within the same programming platform.

\section*{Methods}
\begin{enumerate}[leftmargin=0pt, labelindent=0pt]
\item[] \textbf{Sample preparation.} Single-crystalline samples of 2D bulk WSe$_2$ (2\textit{H} stacking) were purchased from HQ Graphene. Crystals of size around 5 mm $\times$ 5 mm $\times$ 1 mm were used directly for the measurements. To prepare a clean surface by cleaving, we attached a cleaving pin upright to the sample surface using conducting epoxy (EPOTEC H20) outside the vacuum chamber and removed the pin by mechanical force in ultrahigh vacuum.
    
\item[] \textbf{Photoemission experiments.} The measurements were conducted using the HEXTOF instrument \cite{Kutnyakhov2020} at the DESY FLASH PG-2 beamline \cite{Gerasimova2011} with the free-electron laser (FEL) as well as a laboratory source \cite{Puppin2018} with a METIS electron momentum microscope (SPECS METIS 1000) installed at the FHI. In the measurements at FLASH, the FEL was tuned to 36.5 eV (or 34.0 nm) and 109 eV (or 11.4 nm), the optical pump pulse had a center wavelength of 775 nm. The measurements at the FHI used a 21.7 eV home-built extreme UV source based on high harmonic generation in Ar gas driven by an optical parametric chirped-pulse amplifier operating at 500 kHz repetition rate \cite{Puppin2015}. The optical pump pulse is centered at 800 nm. In both FEL and laboratory experiments, the near-infrared light pulses promote the electronic population at the K and K$^\prime$ high-symmetry points (corresponding to $\overline{\mathrm{K}}$ and $\overline{\mathrm{K}^{\prime}}$ points, respectively, in the projected Brillouin zone obtained from photoemission, as shown in Fig. \ref{fig:datarepr}) in momentum space to the excited states via direct optical transitions. The nonequilibrium electronic dynamics are probed via valence and conduction band photoemission \cite{Bertoni2016} as well as core-level photoemission \cite{Dendzik2020}, using $s$-polarized extreme UV and soft X-ray probe pulses, respectively.

\item[] \textbf{Digitization artifact.} The time-to-digital converter (TDC) outputs digitized data according to the binning width of the on-board electronics. Data conversion from one digitized format to another in a rebinning process often creates a picket fence-like effect (see Fig. \ref{fig:jitter}). This phenomenon originates from the incommensurate bin size in the two rounds of sampling processes (binning and rebinning). To solve the problem, one introduces a slight amount of uniformly distributed noise, with an amplitude equal to half of the original bin size, to the single-event values when carrying out the bin counts. This is similar to the histogram jittering (or dithering) technique \cite{Chambers1983,Novo2008} used in statistical visualization and computer graphics. Mathematically, the uniformly distributed noise $U(0, 1)$ bounded in the range $[0, 1]$ is added before binning a univariate data stream, $S=\{ S_i \}$ via,
\begin{equation}
    S_i^{\prime} = S_i + \frac{w_b}{2} \times U(0, 1).
\end{equation}
Here, $w_b$ is the bin width. For binning of multivariate data streams, such as the detector X position (or $k_x$), Y position (or $k_y$), and the photoelectron TOF (or $E$), we adopt the same approach individually for each dimension. The effect of jittering in reducing the digitization artifact is demonstrated in Fig. \ref{fig:jitter}.

\item[] \textbf{Spherical timing aberration.} Electrons entering the TOF tube at different lateral positions travel through different path lengths to reach the detector, which is the origin of the spherical timing aberration as illustrated in Fig. \ref{fig:spherabb}. The lateral position-dependent time delay may be expressed as,
\begin{equation}
  \Delta \text{TOF}_\text{sph}(r)=\big(\sqrt{1 + r^2/d^2} - 1 \big) \text{TOF}_0,
  \label{eq1}
\end{equation}
where $r$ is the radial distance from the center of the DLD and $\text{TOF}_0$ is the TOF normalization constant. For a typical field-free region length of $d \sim 1$ m in the TOF tube and a DLD screen radius of $r=50$ mm, $\Delta \text{TOF}/\text{TOF}_0\approx1.25\times10^{-3}$. Assuming $\text{TOF}_0=0.5$ $\mu s$, the spherical timing aberration in TOF scale is $\Delta \text{TOF}_{\text{sph}}\approx0.62$ ns, which is larger than the DLD's temporal resolution of $\sim$ 0.15 ns. The effect of the spherical timing aberration is visible for a few eV energy range with fine bins but quite small on a large energy range. To illustrate this effect, we use the W$4f$ core-level data presented in Fig. \ref{fig:spherabb}b. For every $(X,Y)$ position on the detector the peak of W$4f_{7/2}$ was fitted with a Voigt profile and the peak positions are shown in Fig. \ref{fig:spherabb}c. As the spectra from deep core levels typically do not show dispersion, the deviation from fitting corresponds to the spherical timing aberration of the electron optics. In order to compensate for the spherical timing aberration, we first transform the data from Cartesian to the polar coordinates (see Fig. \ref{fig:spherabb}c), and then fit the radial-averaged peak position to a polynomial function of the radius,
\begin{equation}
  \Delta \text{TOF}_\text{sph}(r) = \frac{r^2\text{TOF}_0}{2d^2} - \frac{r^4\text{TOF}_0}{8d^4} + \text{O}(r^6).
  \label{eq2}
\end{equation}
The fitting results together with the corrected radial distribution are presented in Fig. \ref{fig:spherabb}d.

\item[] \textbf{Symmetry distortion.} Photoemission patterns in the $(k_x, k_y)$ plane (i.e.~an energy slice) may exhibit distorted symmetry due to the influence of various factors from the instrument, the sample and the experimental geometry on the trajectory of low-energy photoelectrons. Correction of the symmetry distortion yet preserving the intensity features requires the use of symmetry-related landmarks to solve for the symmetrization coordinate transform in the framework of nonrigid image registration \cite{Xian2019}. In typical situations with an excellent electron lens alignment, the energy dependence of the momentum distortion within the focused phase space volume covering an energy range of several eV is negligible, so the same coordinate transform can be applied to all energy slices in the volumetric data (including both valence and conduction bands) or simultaneously to all single events.

\item[] \textbf{Other single-experiment artifacts.} (1) Momentum center shift: The momentum center of the emergent photoelectrons travelling through the electron-optic system may experience an energy-dependent shift owing to the slight misalignment in the system or the influence of stray fields. Correction of the center shift requires an energy-dependent center alignment of energy slices. The shift along the energy (or TOF) axis may be estimated using phase correlation \cite{Guizar-Sicairos2008} or mutual information-based \cite{P.Viola1997} sequential image registration methods, in which the series of energy slices are treated as an image sequence. In a well-shielded and well-aligned electron-optic lens system, generally, the momentum center shift is negligible in the focused photoelectron energy range. (2) Space-charge effect (SCE): The secondary photoelectron clouds originating from the probe and pump pulses cause a ``doming effect" of the photoemission intensity distribution around the momentum center of the band structure. This is especially visible in systems with a clear Fermi edge \cite{Schonhense2015,Schonhense2018} or non-dispersing shallow core levels, which may be used as references for calibrating the parameters used for the flattening transform by applying a momentum-dependent shift $\Delta \text{TOF}_{\text{sc}}(k_x, k_y)$ in the TOF (or the calibrated energy) coordinate of the single-event data.

\item[] \textbf{Momentum calibration.} The scaling factors for momentum calibration are computed by comparing the positions of known high symmetry points in the band structure with their corresponding locations in an energy slice. Suppose $A$ and $B$ are two high symmetry points identifiable (e.g.~as local extrema) from the experimental data with pixel positions ($X_A$, $Y_A$) and ($X_B$, $Y_B$), and momentum positions, ($k_x^A$, $k_y^A$) and ($k_x^B$, $k_y^B$), respectively. We calculate the pixel-to-momentum scaling ratios, $f_X$ and $f_Y$, along the $X$ (column) and $Y$ (row) directions of a 2D $k$-space image, respectively. Then, the momentum coordinate ($k_x$, $k_y$) at each pixel position ($X$, $Y$) may be derived.
\begin{align}
    f_D &= (k_d^A - k_d^B) / (D_A - D_B) \\
    k_d &= f_D \times (D - D_A) \quad (D, d = X, x \text{ or } Y, y)
\end{align}

\item[] \textbf{Energy calibration.} The calibration requires a set of band mapping data measured at different bias voltages (applied between the material sample and the ground), usually sampled with a spacing of 0.5 V in a range of $\pm$3-5 V around the normally applied bias voltage for a particular sample. The calibration proceeds by finding the TOF feature (e.g. local extrema) correspondences in the 1D energy distribution curves (EDCs) at different biases using the dynamic time warping algorithm \cite{Salvador2007}. The transformation from the TOF to the photoelectron energy $E$ is approximated as a polynomial function,
\begin{equation}
	E(\text{TOF}) = \sum_{i=0}^n a_i \text{TOF}^i
\end{equation}
The approximation is sufficiently accurate within a range of $\sim$ 20 eV, sufficient to cover the entire valence band and some low-lying conduction bands of typical materials. The polynomial coefficients are determined using nonlinear least squares by solving $\Delta T \cdot\textbf{a}=\Delta E$, in which $\textbf{a} = (a_1, a_2, ...)^T$ is the coefficient vector while the constant offset $a_0$ is determined by manual alignment to an energy reference, such as the Fermi level or valence band maximum. The vector $\Delta E$ and the matrix $\Delta T$ contain, respectively, the pairwise differences of the bias voltages and the polynomial terms of differential $\text{TOF}$ values. To calibrate a large energy range including multiple core levels, a piecewise polynomial may be used \cite{Schonhense2018}.

\item[] \textbf{Pump-probe delay calibration.} The time origin (``time zero") in time-resolved photoemission spectroscopy, i.e.~the temporal overlap of pump and probe pulses, is determined by fitting of a characteristic trace extracted from the data. Since the readings of the digital encoder (see Fig. \ref{fig:workflow_mapping}) are sampled linearly, equally-spaced pump-probe delays are directly convertible from the readings using linear interpolation, given the boundary values of the translation stage positions and the corresponding delay times. For unequally-spaced delays, a delay marker is first added to each data point as a separate column after data acquisition to group together the encoder reading ranges that correspond to the specific time delays. The data binning is carried out over the delay marker column instead of the equally-sampled encoder readings.

\item[] \textbf{Visualization strategies.} We discuss here three methods for the display of volumetric band mapping data, which are, at the same time, the basis for visualizing 3D+t data with time as an animated axis. (1) The orthoslice representation includes orthogonal 2D planes selected in specific regions in the 3D volume \cite{Hansen2005}, which highlights specific slices deep within the data less visible in a volumetrically rendered view (see Fig. \ref{fig:datarepr}a). Along this line, we have developed a software package, \textsf{4Dview} \cite{Dendzik2019}, to explore 4D data using simultaneously linked orthoslices, which also features contrast adjustment and data integration within a hypervolume of interest. (2) The band-path plot (see Fig. \ref{fig:datarepr}b) is a 2D representation of the 3D band mapping volume generated by combining a series of 2D cuts along selected momentum paths (or k-paths) traversing a list of so-called high-symmetry points \cite{Setyawan2010,Hinuma2017}. This representation captures the largest dispersion within the band structure. For volumetric data, the same path may be sampled from all the full energy range to produce the plot shown in Fig. \ref{fig:datarepr}b. The \textsf{analysis} and \textsf{visualization} modules in the \textsf{mpes} package include functionalities to compose customized band-path plots. (3) The cut-out view (see Fig. \ref{fig:datarepr}c) exposes a specific part of interest in the volumetric data, while not losing the rest \cite{Hansen2005}. The \textsf{analysis} module in the \textsf{mpes} package provides ways to generate precise cut-outs using position landmarks (e.g. high-symmetry points labelled in Fig. \ref{fig:datarepr}) and inequalities.
\end{enumerate}

\section*{Acknowledgements}
We thank G.~Sch{\"o}nhense for support on the photoelectron detector, S.~Grunewald, S.~Sch{\"u}lke and G.~Schnapka for support on the computing infrastructures. We thank G.~Brenner, H.~Redlin and S.~Dziarzhytski at FLASH, DESY, and H. Meyer and S. Gieschen from the University of Hamburg for beamline and instrumentation support. The work was partially supported by BiGmax, the Max Planck Society's Research Network on Big-Data-Driven Materials-Science, the European Research Council (ERC) under the European Union's Horizon 2020 research and innovation program (Grant No. ERC-2015-CoG-682843), and the German Research Foundation (DFG) through the Emmy Noether program under grant number RE 3977/1 and the
SFB/TRR 227 ”Ultrafast Spin Dynamics” (projects A09 and B07). D. Kutnyakhov, M. Heber and W. Wurth acknowledge funding by the DFG within the framework of the Collaborative Research Centre SFB 925 - 170620586 (project B2). F. Pressacco acknowledges funding from the excellence cluster EXC 1074 “The Hamburg Centre for Ultrafast Imaging - Structure, Dynamics and Control of Matter at the Atomic Scale” of the DFG. S. Y. Agustsson and J. Demsar acknowledge the financial support by the DFG in the framework of the Collaborative Research Centre SFB TRR 173 - 268565370  (project A5). D. Curcio and P. Hofmann acknowledge funding from VILLUM FONDEN via the Centre of Excellence for Dirac Materials (Grant No. 11744). T. Pincelli thanks the Alexander von Humboldt Foundation for financial support.

\section*{Author contributions}
Y.A., K.B., S.Y.A., D.C., R.P.X. and M.D. wrote the \textsf{hextof-processor} package. R.P.X. and L.R. wrote the \textsf{mpes} package. D.K., Y.A., F.P., R.P.X., S.Y.A., D.C., M.D., M.H., S.D., P.H., L.R., R.E.~and W.W.~participated in the experiments at the FLASH PG-2 beamline using the HEXTOF instrument in Hamburg. S.D.~and L.R.~conducted the experiment at the Fritz Haber Institute using the METIS electron momentum microscope. R.P.X., M.D., L.R., R.E., M.S., T.P. constructed the metadata format, R.P.X. and M.S. implemented them into \textsf{parser-mpes}. R.P.X. wrote the initial manuscript with contributions from M.D. and Y.A. All authors contributed to discussions to bring the manuscript to its final form.

\section*{Data availability}
The single-event photoemission data used for demonstrating the workflow is available on the Zenodo platform at \href{https://doi.org/10.5281/zenodo.2704787}{10.5281/zenodo.2704787} (valence and conduction band photoemission at FEL) \cite{Xian2020dataa}, \href{https://doi.org/10.5281/zenodo.3945432}{10.5281/zenodo.3945432} (core-level photoemission at FEL) \cite{Dendzik2020data} and \href{https://doi.org/10.5281/zenodo.3987303}{10.5281/zenodo.3987303} (valence band photoemission from laboratory setup). \cite{Xian2020datab} The preprocessed data are being integrated into the NOMAD database in the domain for experimental materials science data accessible at \href{https://nomad-lab.eu/prod/rae/gui/search?domain=ems}{https://nomad-lab.eu/prod/rae/gui/search?domain=ems}.

\section*{Code availability}
The code, including documentation and examples in Jupyter notebooks for implementing the data transformations in the workflow are available as \textsf{hextof-processor} \\(\href{https://github.com/momentoscope/hextof-processor}{https://github.com/momentoscope/hextof-processor}) \cite{hextof-processor} and \textsf{mpes} (\href{https://github.com/mpes-kit/mpes}{https://github.com/mpes-kit/mpes}) \cite{mpes}. The parser for integrating preprocessed experimental data into the NOMAD database is available as \textsf{parser-mpes} (\href{https://gitlab.mpcdf.mpg.de/rpx/parser-mpes}{https://gitlab.mpcdf.mpg.de/rpx/parser-mpes}) \cite{parser-mpes}.

\section*{Competing interests}
The authors declare no competing interests.

\bibliographystyle{naturemag}

\end{document}